**An improved phase-field framework for simulating impacts of solidifying metal drops**

**Ali Mostafavi[a], Vitaliy Yurkiv[a], Alexander L. Yarin[b], Farzad Mashayek[a,*]**

[a] *Department of Aerospace and Mechanical Engineering, University of Arizona, Tucson, AZ 85721, USA*

[b] *Department of Mechanical and Industrial Engineering, University of Illinois Chicago, Chicago, IL 60607, USA*

Here, an improved phase-field method for simulating the impact dynamics of solidifying molten metal droplets is developed using targeted free-energy modifications. Conventional Cahn–Hilliard–Navier–Stokes (CHNS) formulations generally do not capture melt retraction over a solidified portion of the droplet, because the newly formed solid region is not treated as an actual internal boundary in a single-order-parameter diffuse-interface model. As a result, the formulation lacks an internal wetting condition or localized wall-energy mechanism capable of driving melt retraction over a solidified splat. To address this limitation, a diffuse-domain wall-energy term is added to the free energy, with a chemical-potential contribution that is active near the diffuse liquid–solidified-material–gas triple-line region, enabling the remaining melt to retract over a solidified splat. In addition, a solidification penalty term is introduced to immobilize the solidified splat formed during impact and suppress unphysical interface motion caused by residual Cahn–Hilliard diffusion inside the frozen region. The proposed formulation is validated against benchmark experiments on impacts of solidifying tin droplets. The results reveal that the localized wall-energy term captures post-maximum-spread melt retraction, while the penalty term effectively arrests motion of the solidified splat. Qualitative and quantitative comparisons with experiments, volume-of-fluid simulations, and standard phase-field predictions demonstrate that the proposed formulation captures post-maximum-spread melt retraction and provides an accurate estimate of the stabilized final splat diameter.

## 1. Introduction

Thermal spray coating is a coating-manufacturing process in which metal or ceramic powders are injected into a high-temperature gas jet, where they are melted and propelled toward the substrate to form a coating [1]. Upon impact, the molten droplets spread rapidly, solidify, and form individual splats that collectively build up a coating. The mechanical properties of the final coating, including porosity, microstructure, and adhesion strength, are strongly influenced by the morphology of such splats, which is governed by the coupled impact, spreading, and solidification dynamics of individual droplets and droplet–droplet interactions [2]. Therefore, coating quality depends on many process parameters, including the number of droplets, droplet size distribution, impact velocity, temperature, solidification rate, and thermal contact resistance. Achieving better control over the process outcome requires fundamental understanding of the fluid mechanics and heat transfer processes involved in the impacts of the individual solidifying droplets. Significant analytical and experimental results pertaining impacts of solidifying droplets are discussed in [3].

Various numerical techniques have been developed to model free-surface flows undergoing phase change. Earliest numerical simulations of impacts of solidifying metal drops employed the volume-of-fluid (VOF) method. Using the commercial code FLOW-3D, Trapaga et al. [4] modeled spreading and solidification of molten metal droplets on solid substrates and validated their predictions against high-speed experimental measurements. Another VOF-based study by Liu et al. [5] used the RIPPLE code to simulate droplet impact and freezing, applying a two-phase continuum model for the flow with a growing solid layer. Pasandideh-Fard et al. [2] used a modified SOLA-VOF method for 2D axisymmetric simulations of molten tin droplet impact on stainless steel substrates, with both contact angle and thermal contact resistance determined experimentally and prescribed in the numerical model. Later, Pasandideh-Fard et al. [6] extended the VOF framework to a full 3D model for normal and oblique impacts of molten tin droplets on stainless steel substrates, with the normal-impact simulations validated against the experiments of Aziz and Chandra [1]. More recently, Zhang et al. used the VOF method implemented in ANSYS Fluent in conjunction with an enthalpy-porosity solidification model to simulate sessile water droplet freezing, including conical-tip formation, and impacting-freezing dynamics of supercooled water droplets on cold superhydrophobic flat surfaces [7,8].

* Corresponding author.
*E-mail address*: mashayek@arizona.edu (F. Mashayek).

Compared with the VOF method, phase-field methods avoid explicit interface reconstruction and allow surface tension and wetting effects to be introduced naturally through a free-energy formulation. Jacqmin laid the groundwork for phase-field modeling of isothermal free-surface flows by using a single order parameter governed by a conservative Cahn–Hilliard equation and coupling it to the Navier–Stokes equations through a free-energy-based surface-tension force [9]. In addition, he showed how the wall-free-energy term at the solid boundary can be used to impose wetting effects in diffuse-interface formulations [10].

Solidifying molten drops are inherently three-phase flows involving liquid, solid, and gas phases. Wang et al. developed an energy-stable two-order-parameter phase-field model, in which a non-conservative Allen–Cahn equation describes liquid–solid phase change and a conservative Cahn–Hilliard equation captures the liquid–gas interface [11]. Their model satisfies an energy-dissipation law and captures the pointy-tip formation of frozen sessile droplets, revealing better agreement with experiments than their corresponding ANSYS Fluent VOF predictions. Despite the advantages, using multiple order parameters introduces additional degrees of freedom into the nonlinear system of equations and requires extra model parameters associated with the additional phase-field equation.

Shen et al. simulated the normal impact of solidifying droplets using a modified Cahn–Hilliard–Navier–Stokes (CHNS) formulation tailored to droplet impact with solidification [12]. Their formulation uses a single order parameter to distinguish the condensed phase from the gas phase, while the liquid and solid regions within the condensed phase are identified through a temperature-dependent liquid fraction. This liquid fraction varies smoothly across a narrow artificial mushy region, allowing the solidification front to be captured implicitly within the phase-field framework. Compared to the standard CHNS formulation, they added a Kozeny–Carman-type momentum sink term to the momentum balance equation to damp velocity inside the solidified portion of the droplet. Their numerical results for solidifying tin drops are compared with the experiments of Aziz and Chandra [1]. While a reasonable agreement with the experiments is obtained from the moment of impact until the maximum droplet spread, the model does not fully capture some important post-maximum-spread dynamics, such as melt retraction over its own solidified portion.

Physically, while velocity-suppression treatments can immobilize the solidified region by driving its velocity toward zero, the remaining melt does not experience itself as sitting atop a solid wall comparable to the substrate. Within the phase-field framework, the solidified region is an evolving part of the computational domain rather than a fixed geometric boundary. As a result, imposing a geometric contact-angle condition along the evolving liquid–solidified-splat–gas region remains an open modeling challenge within this numerical framework.

In the present work, inspired by wall-energy treatment at the liquid–solid contact line, this limitation is addressed by incorporating a diffuse-domain wall-energy term into the free energy of the standard CHNS formulation [13,14]. The effect of this term is localized near the diffuse liquid–solidified-material–gas triple-line region and provides an additional energy-driven mechanism that enables the remaining melt to retract or spread over its solidified portion. In addition, Cahn–Hilliard diffusion is known to cause residual interface motion even in regions where fluid motion is suppressed [15]. This becomes problematic near the solidified drop–gas interface, where the solid velocity is arrested, but the phase-field variable can still drift due to the Cahn–Hilliard diffusion. To suppress this artifact, a solidification penalty term is incorporated into the free-energy functional, immobilizing the phase field inside the solidified region without imposing artificial pinning.

The article is organized as follows. Section 2 presents the governing equations for heat transfer, the proposed phase-field formulation including the wall-energy and solidification-penalty terms, and fluid flow. Section 3 describes the numerical method, including the finite-element implementation, boundary and initial conditions, and adaptive mesh refinement strategy. Section 4 presents the benchmark one-dimensional Stefan problem and isothermal droplet impact as validation cases, followed by the simulation of solidifying molten tin droplet impact. Finally, Section 5 summarizes the main findings and draws conclusions of the study.

## 2. Governing equations

The impact dynamics of solidifying molten metal droplets on a solid surface are governed by the combined effects of fluid flow, wetting, heat transfer, and phase change. As shown in Fig. 1, the solidifying droplet undergoes rapid spreading after impact, followed by basal solidification and contact-line arrest on the substrate. After maximum spread, the remaining melt recoils and retracts over the previously solidified portion, eventually forming a final solidified splat whose morphology is governed by the coupled spreading, recoil, and solidification dynamics.

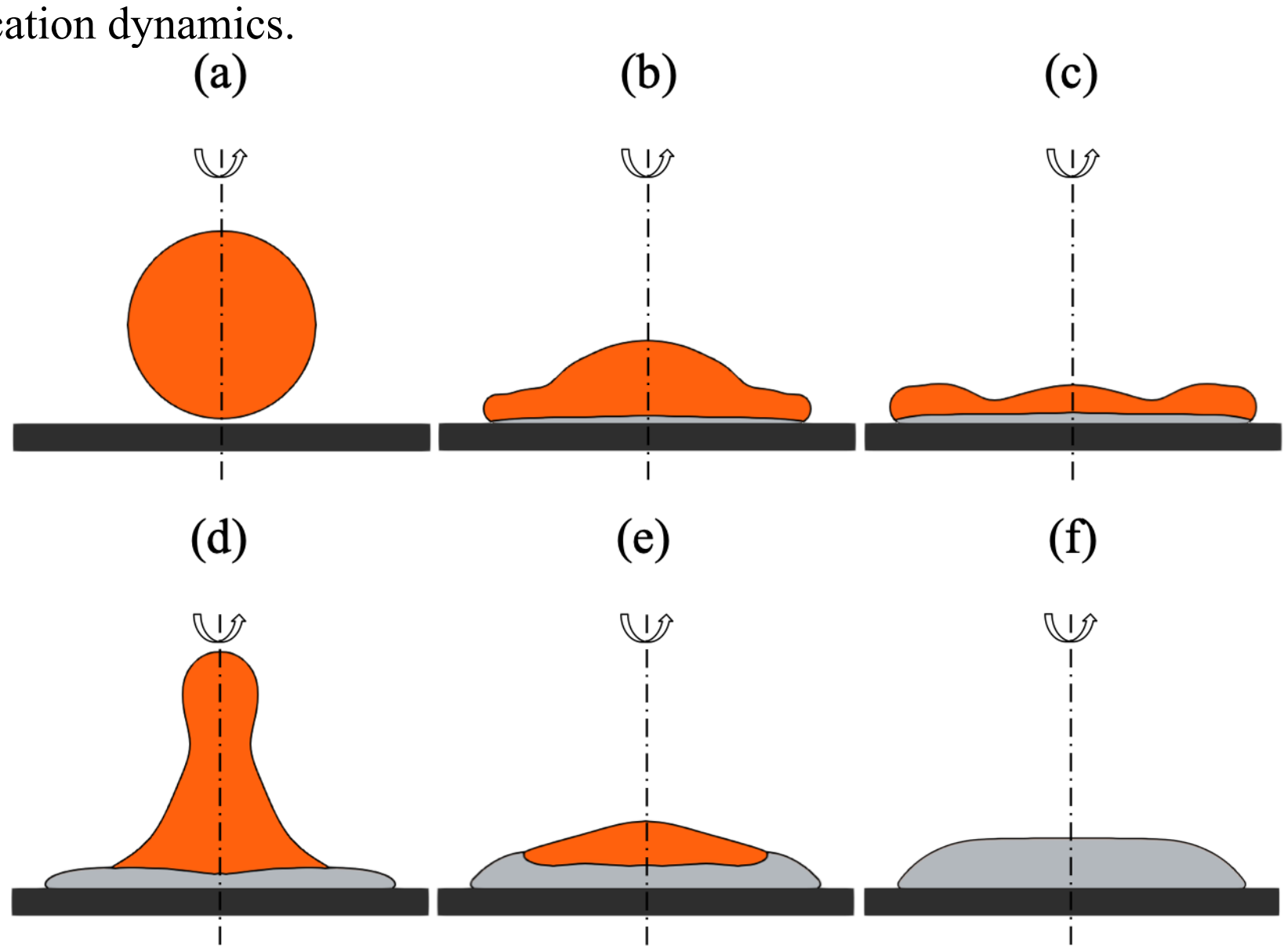

**Fig. 1.** Schematic of an axisymmetric solidifying molten droplet following the normal impact sequence: (a) pre-impact, (b) spreading, (c) maximum spread with contact-line arrest, (d) melt recoil and retraction over the solidified splat, (e) late-stage melt redistribution, and (f) final solidified splat morphology.

The following subsections present the mathematical model used to describe the problem and introduce the proposed free-energy modifications for capturing melt retraction and enforcing immobilization of the solidified region.

### 2.1. *Heat transfer*

Heat transfer inside the solidifying droplet and the surrounding gas is modeled using an enthalpy-based phase-change formulation, with viscous dissipation neglected [12,16,17]:

$$\rho c_p \left( \frac{\partial T}{\partial t} + \boldsymbol{u} \cdot \boldsymbol{\nabla} T \right) = \boldsymbol{\nabla} \cdot (k \boldsymbol{\nabla} T) - \rho L \frac{\partial f_l}{\partial t} G(c), \tag{1}$$

where $\rho$ is the density, $c_p$ is the specific heat, $T$ is the temperature, $\boldsymbol{u}$ is the velocity vector, $k$ is the thermal conductivity, $L$ is the latent heat of fusion, $f_l$ is the liquid fraction defined across the diffuse mushy region, and $c$ is the order parameter distinguishing the condensed phase ($c = 1$) from the gas phase ($c = -1$). The latent heat of fusion released during solidification is incorporated as a source term in the energy equation and is active only within the diffuse mushy region near the solidification front. The function $G(c) = (1 + c)/2$ is a droplet-gate indicator that takes a value of unity in the condensed phase and zero in the gas phase, thereby localizing the latent-heat source to the droplet.

The liquid fraction $f_l$ is defined as a temperature-dependent function across a narrow interval around the melting temperature $T_m$. The solidus and liquidus temperatures are defined as $T_{\mathrm{sol}} = T_m - \delta T/2$ and $T_{\mathrm{liq}} = T_m + \delta T/2$, where $\delta T$ is a small temperature offset used to regularize the solid–liquid transition.

$$f_l(T) = \begin{cases} 0, & T \leq T_{\mathrm{sol}}, \\ 6{\eta_T}^5 - 15{\eta_T}^4 + 10{\eta_T}^3, & T_{\mathrm{sol}} < T < T_{\mathrm{liq}}, \eta_T = \dfrac{T - T_{\mathrm{sol}}}{T_{\mathrm{liq}} - T_{\mathrm{sol}}}, \\ 1, & T \geq T_{\mathrm{liq}}. \end{cases} \tag{2}$$

Inside the condensed phase, $f_l = 0$ represents the solid phase and $f_l = 1$ represents the liquid phase. The quintic smootherstep interpolation provides a smooth transition between the two limits [18].

Heat transfer within the solid substrate is modeled by pure conduction, since no fluid motion occurs in this region:

$$\rho_w c_{p,w} \frac{\partial T_w}{\partial t} = \boldsymbol{\nabla} \cdot (k_w \boldsymbol{\nabla} T_w). \tag{3}$$

Here, $T_w$ is the substrate temperature, while $\rho_w$, $c_{p,w}$, and $k_w$ denote the density, specific heat, and thermal conductivity of the substrate, respectively.

Treating the molten droplet–substrate interface as perfectly thermally coupled is unrealistic [19]. During an impact onto a rough solid surface, incomplete contact and trapped air within surface asperities can form an insulating interfacial layer, whose resistance depends on surface finish, contact pressure, and material properties [19]. This produces a temperature jump between the droplet and substrate, which is represented by a thermal contact resistance $R_c$,

$$R_c = \frac{T_d - T_w}{q''}, \tag{4}$$

where $q''$ is the heat flux from the droplet to the substrate. The thermal contact resistance $R_c$ is prescribed as a model input. Although $R_c$ may vary spatially and temporally during droplet impact [20], a constant value is assumed in the present simulations.

### 2.2. *Proposed solidification-aware phase-field formulation*

The advective Cahn–Hilliard equation is used to describe the evolution of the free surface [21],

$$\frac{\partial c}{\partial t} + \boldsymbol{u} \cdot \boldsymbol{\nabla} c = \boldsymbol{\nabla} \cdot (M \boldsymbol{\nabla} \psi), \tag{5}$$

where $M$ is the phenomenological mobility, and $\psi$ is the chemical potential.

The chemical potential $\psi$ is derived from a single-order-parameter Ginzburg-Landau free-energy functional $\mathcal{F}[c, \boldsymbol{\nabla} c]$ defined over the computational domain $\Omega$ [22]. For solidifying droplets, this functional is constructed as the sum of four contributions:

$$\mathcal{F}[c, \nabla c] = \int_{\Omega} \left(f_b + f_m + f_w + f_p\right) d\Omega. \tag{6}$$

The four free-energy contributions in Eq. (6) are described below.

2.2.1. *Classical Cahn–Hilliard formulation*

The first contribution, $f_b$, is the bulk double-well free-energy density:

$$f_b = \frac{\sigma}{4\xi^2}(c^2 - 1)^2. \tag{7}$$

This term has minima at $c = \pm 1$ and represents the phobic contribution to the free energy, favoring separation between the condensed phase ($c = 1$) and the gas phase ($c = -1$). In Eq. (7), $\xi$ controls the diffuse-interface thickness, while $\sigma$ denotes the mixing energy density.

The second term, $f_m$, is the gradient-energy contribution:

$$f_m = \frac{1}{2}\sigma|\nabla c|^2. \tag{8}$$

This term represents the philic contribution, since it favors smooth mixing across a finite-thickness diffuse interface. The mixing energy density $\sigma$ is related to the interfacial surface-tension coefficient $\gamma$ and diffuse-interface thickness $\xi$ through [23,24]

$$\sigma = \frac{3}{2\sqrt{2}}\gamma\xi. \tag{9}$$

2.2.2. *Diffuse-domain wall-energy term*

The third contribution, $f_w$, is introduced as a diffuse-domain wall-energy density to mimic the wetting effect that would normally be imposed through a contact-angle condition at a solid boundary [13,14]. In the classical phase-field contact-line formulation, the wall-energy contribution is defined on the solid boundary and contains the factor $\gamma\cos\theta$, where $\gamma$ is the gas–liquid surface-tension coefficient and $\theta$ is the prescribed contact angle. In the present single-order-parameter formulation, however, the solidified portion of the droplet is not represented as an explicit internal boundary. Instead, it is described by a diffuse thermal solidification indicator. As a result, the interaction between the remaining melt, gas, and solidified material occurs over a finite diffuse region rather than at a sharp contact line. A geometric contact angle is therefore not prescribed explicitly. The classical factor $\cos\theta$ is replaced by an effective dimensionless coefficient, $\beta_w$, while $\gamma$ is retained to preserve the physical surface-energy scale.

The diffuse-domain wall-energy density is written as

$$f_w = \chi_r(t)\beta_w\delta_{sf}(\mathbf{x}, t)g_w(c), \tag{10}$$

where $\chi_r(t)$ is an activation function that is zero before maximum spreading and becomes unity once the contact-line velocity changes sign, $\beta_w$ controls the strength and sign of the effective wetting or dewetting bias, and $\delta_{sf}(\mathbf{x}, t)$ localizes the contribution near the solidification-front region. The wall-energy function $g_w(c)$ is defined as [14]

$$g_w(c) = -\gamma\frac{c(3 - c^2)}{4}. \tag{11}$$

The solidification-front localization function is given by

$$\delta_{sf}(\mathbf{x}, t) = |\nabla f_l|, \tag{12}$$

which is nonzero only across the diffuse thermal solidification front. Since $f_l$ is dimensionless, $\delta_{sf}(\mathbf{x}, t)$ has units of the inverse length. The wall-energy function $g_w(c)$ has units of energy per unit area because it is scaled by the surface tension $\gamma$. Therefore, $f_w$ has units of energy per unit volume.

2.2.3. *Solidification-penalty term*

The fourth contribution, $f_p$, is introduced as a solidification penalty energy density to immobilize the phase-field variable in regions that have solidified. In the standard Cahn–Hilliard formulation, order-parameter diffusion remains active even after the local material becomes solid. Thus, although the velocity in the solidified region may be suppressed, a residual Cahn–Hilliard diffusion can still cause artificial drift of the gas–solid interface [15]. To suppress this artifact, $c$ is penalized toward a stored frozen target value once local solidification is detected, thereby weakly anchoring the phase field inside the solidified region:

$$f_p = \frac{1}{2}\mathrm{K}_f(\mathbf{x}, t)\left(c - c_{\text{target}}\right)^2. \tag{13}$$

Here, $c_{\text{target}}$ is the stored frozen target value and $\mathrm{K}_f(\mathbf{x}, t)$ is the penalty coefficient, defined as

$$\mathrm{K}_f(\mathbf{x}, t) = \mathrm{K}_0 S_f(\mathbf{x}, t), \tag{14}$$

where $\mathrm{K}_0$ is the penalty strength and $S_f$ is a binary freeze mask defined as:

$$S_f^{n+1} = (1 - A_m^{n+1}) \max\left[S_f^n,\ \mathcal{I}\left(f_l(T^n) \le f_{l,\text{trig}}\right)\mathcal{I}\left(c^n \ge c_{\text{trig}}\right)\right]. \tag{15}$$

The superscript $n$ denotes the discrete time level, with $t^{n+1} = t^n + \Delta t$, and $\mathcal{I}(\cdot)$ denotes the indicator function, which equals unity when the condition inside the parentheses is satisfied and zero otherwise. The freeze mask $S_f^{n+1}$ is updated from its previous value $S_f^n$, so that once a material element becomes frozen, it remains frozen unless melting is allowed through the remelting indicator $A_m$. For irreversible solidification, $A_m = 0$; hence, once activated, $S_f$ remains a latched freeze mask. Since the liquid fraction satisfies $f_l = 0$ in the solidified region and $f_l = 1$ in the liquid region, freezing is detected when $f_l(T^n) \le f_{l,\text{trig}}$, where $f_{l,\text{trig}}$ is a prescribed freezing threshold. The condensed-phase indicator, $\mathcal{I}\left(c^n \ge c_{\text{trig}}\right)$, prevents the freeze mask from activating in the gas phase by restricting freezing to regions where the phase field corresponds to the condensed phase.

The stored frozen target value is updated as

$$c_{\text{target}}^{n+1} = \begin{cases} c_s, & S_f^n = 0 \text{ and } S_f^{n+1} = 1, \\ c_{\text{target}}^n, & S_f^n = 1 \text{ and } S_f^{n+1} = 1, \\ c^n, & S_f^{n+1} = 0, \end{cases} \tag{16}$$

where $c_s$ is the prescribed solid-state target value, taken as $c_s = 1$. Thus, when a material element freezes for the first time, the phase-field variable is anchored toward $c_s$. If the material element remains frozen, the previous target value is retained. If the material element is not frozen, the penalty coefficient vanishes because $S_f = 0$, and the stored value does not affect the solution. This formulation suppresses artificial phase-field drift inside solidified regions while leaving unfrozen and gas-phase regions unaffected, without prescribing a separate contact-line pinning condition after maximum spreading.

### 2.2.4. *Resulting chemical potential*

Having defined the four contributions to the total free-energy functional, the phase-field chemical potential is obtained by taking the variational derivative of the total free energy with respect to the order parameter $c$, such that $\psi = \delta\mathcal{F}/\delta c$:

$$\psi = \sigma\left(\frac{c^3 - c}{\xi^2} - \nabla^2 c\right) + \chi_r(t)\beta_w\delta_{sf}(\mathbf{x},t)g_w'(c) + \mathrm{K}_f(\mathbf{x},t)\left(c - c_{\text{target}}\right). \tag{17}$$

Although the wall-energy density $f_w$ has nonzero values in the pure phases ($c = \pm 1$), the quantity that enters the chemical potential equation is its derivative. The corresponding wall-induced chemical-potential contribution is proportional to $g_w'(c) = dg_w/dc$, which vanishes at $c = \pm 1$. Therefore, this contribution does not force the phase field in the pure gas or pure condensed phase. As illustrated in Fig. 2, it is active only where the thermal solidification-front localization $\delta_{sf}(\mathbf{x},t)$ overlaps the diffuse gas–condensed-phase interface. This localized overlap corresponds to the diffuse liquid–frozen-splat–gas triple-line region, where the remaining melt can retract over the already solidified splat.

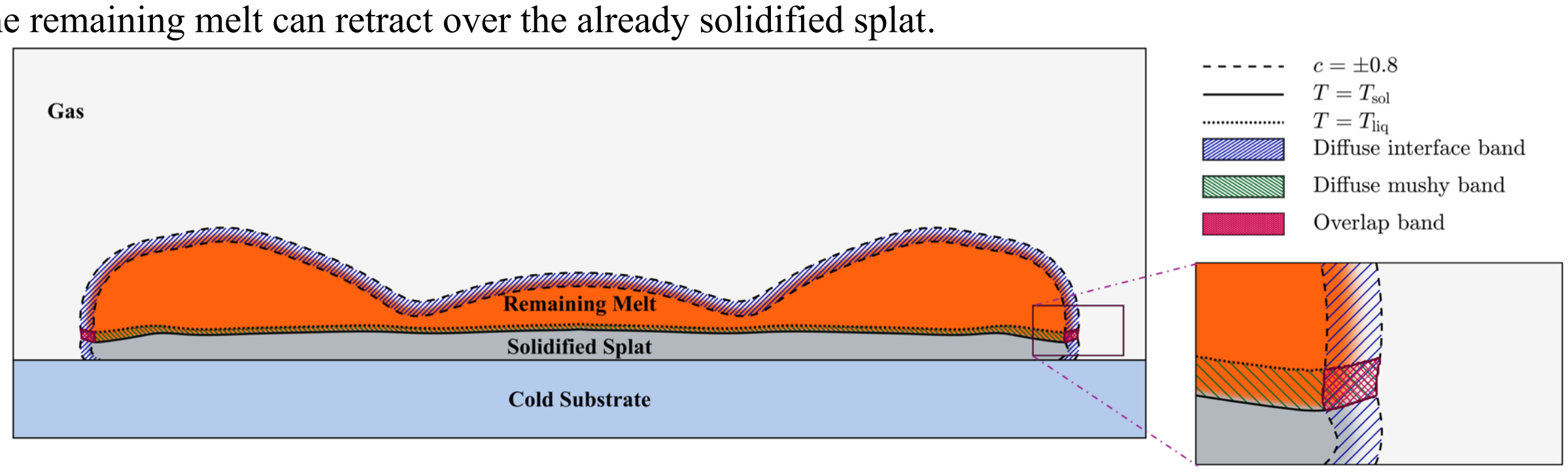

**Fig. 2.** Localization of the wall-induced chemical-potential contribution near the diffuse liquid–frozen-splat–gas triple-line region. The dashed contours denote the diffuse gas–condensed-phase interface, represented by $c = \pm 0.8$. The solid and dotted temperature contours denote $T_{\text{sol}}$ and $T_{\text{liq}}$, respectively, marking the lower and upper limits of the thermal mushy region. The highlighted overlap between the diffuse interface band and the mushy band identifies the region where the wall-energy contribution is active. The zoomed inset illustrates that this contribution is confined to the neighborhood of the contact region between the remaining melt, the frozen splat, and the surrounding gas.

## 2.3. *Hydrodynamics*

Assuming laminar incompressible flow and neglecting volume changes associated with solidification, the continuity and momentum balance equations are solved to obtain the pressure and velocity fields:

$$\nabla \cdot \boldsymbol{u} = 0, \tag{18}$$

$$\rho\left(\frac{\partial \boldsymbol{u}}{\partial t} + \boldsymbol{u}\cdot\nabla\boldsymbol{u}\right) = -\nabla p + \nabla\cdot\boldsymbol{\tau}_v + \psi\nabla c + \rho\boldsymbol{g}, \tag{19}$$

where $p$ is the modified pressure, $\boldsymbol{\tau}_v = \mu(\nabla\boldsymbol{u} + \nabla\boldsymbol{u}^T)$ is the deviatoric viscous stress tensor, $\mu$ is the viscosity, $\boldsymbol{g}$ denotes gravitational acceleration vector, and the term $\psi\nabla c$ is the volume-spread capillary force [25].

Several approaches can be used to arrest fluid motion in the solidified portion of an impacting droplet. Pasandideh-Fard et al. used a modified fixed-velocity treatment, in which the solidified material was assigned zero velocity and treated as a liquid with effectively infinite density [6]. Another approach is to add a Kozeny–Carman-type momentum sink term, which penalizes the velocity toward zero in the solidified region as the liquid fraction approaches zero [7,12,26,27]. In the present work, the solidified region is instead modeled as a liquid with effectively infinite viscosity, thereby suppressing motion without introducing an additional penalty term [11,16,28].

The density, viscosity, thermal conductivity, and specific heat are interpolated using the liquid fraction within the condensed phase and a phase indicator across the diffuse interface. For a generic material property $X \in \{\rho, \mu, k, c_p\}$, this blend is written as

$$X(c,T) = [1 - G(c)]X_g + G(c)[X_s + (X_l - X_s)f_l(T)], \tag{20}$$

where $X_g$, $X_l$, and $X_s$ denote the gas, liquid, and solid values of $X$, respectively.

## 3. Numerical methods

### 3.1. *Finite element formulation*

The governing equations described in Section 2 are implemented within the Multiphysics Object-Oriented Simulation Environment (MOOSE) finite-element framework [24,29–31]. A Taylor–Hood element pair is used for spatial discretization, with the velocity field approximated using second-order shape functions and all other nonlinear variables interpolated using first-order shape functions. Time integration is performed using a fully implicit second-order backward differentiation formula (BDF2). The weak-form implementation in the fluid domain is described below:

$$-(\boldsymbol{\nabla} \cdot \boldsymbol{u}, \hat{p})_\Omega = 0, \tag{21}$$

$$\left(\rho \frac{\mathrm{D}\boldsymbol{u}}{\mathrm{D}t}, \hat{\boldsymbol{u}}\right)_\Omega - (\psi \boldsymbol{\nabla} c + \rho \boldsymbol{g}, \hat{\boldsymbol{u}})_\Omega + (-p\boldsymbol{I} + \boldsymbol{\tau}_v, \boldsymbol{\nabla}\hat{\boldsymbol{u}})_\Omega - (\boldsymbol{n} \cdot (-p\boldsymbol{I} + \boldsymbol{\tau}_v), \hat{\boldsymbol{u}})_\Gamma = 0, \tag{22}$$

$$\left(\rho\left(c_p + LG\frac{df_l}{dT}\right)\frac{\partial T}{\partial t}, \hat{T}\right)_\Omega + \left(\rho c_p \boldsymbol{u} \cdot \boldsymbol{\nabla} T, \hat{T}\right)_\Omega + \left(k\boldsymbol{\nabla} T, \boldsymbol{\nabla}\hat{T}\right)_\Omega - \left(\boldsymbol{n} \cdot (k\boldsymbol{\nabla} T), \hat{T}\right)_\Gamma = 0, \tag{23}$$

$$\left(\frac{\mathrm{D}c}{\mathrm{D}t}, \hat{\psi}\right)_\Omega + \left(M\boldsymbol{\nabla}\psi, \boldsymbol{\nabla}\hat{\psi}\right)_\Omega - \left(\boldsymbol{n} \cdot (M\boldsymbol{\nabla}\psi), \hat{\psi}\right)_\Gamma = 0, \tag{24}$$

$$\left(\psi - \frac{\partial}{\partial c}\left(f_b + f_w + f_p\right), \hat{c}\right)_\Omega - (\sigma\boldsymbol{\nabla} c, \boldsymbol{\nabla}\hat{c})_\Omega + (\boldsymbol{n} \cdot (\sigma\boldsymbol{\nabla} c), \hat{c})_\Gamma = 0. \tag{25}$$

Hatted quantities denote the corresponding test functions. The notation $(\cdot,\cdot)_\Omega$ denotes the $L^2(\Omega)$ inner product over the computational domain, while $(\cdot,\cdot)_\Gamma$ denotes the corresponding boundary inner product over $\Gamma = \partial\Omega$. The vector $\boldsymbol{n}$ is the outward unit normal on $\Gamma$. The material derivative is defined as $\mathrm{D}/\mathrm{D}t = \partial/\partial t + \boldsymbol{u} \cdot \boldsymbol{\nabla}$. All governing equations in the fluid domain are solved in a fully coupled manner. The resulting nonlinear system is treated using Newton's method, and the linearized system at each Newton iteration is solved using the MUltifrontal Massively Parallel sparse direct Solver (MUMPS), which uses LU factorization for large sparse systems.

### 3.2. *Boundary and initial conditions*

The schematic representation of the solidifying molten droplet normal impact problem is shown in Fig. 3, including the pre-impact configuration and the maximum-spread state. Here, $D_0$ is the initial droplet diameter, $U_0$ is the impact velocity, and $T_0$ is the initial droplet temperature, which is above the melting temperature $T_m$. The cold substrate is initially at $T_{0,w} < T_m$.

The droplet-fluid domain has dimensions $7D_0 \times 7D_0$, while heat transfer in the substrate is calculated separately in the solid domain. Note again that thermal contact resistance is imposed over the wetted droplet-substrate footprint, allowing a finite temperature jump across the droplet-substrate thermal contact interface. Symmetry conditions are applied along the axis of symmetry. A no-slip condition is imposed for the velocity at the substrate ($\boldsymbol{u} = 0$), while the far-field boundaries are placed sufficiently far from the impact region and treated using zero-normal-gradient conditions for the velocity field. Adiabatic thermal conditions are applied to the remaining external boundaries. For the order parameter and chemical potential, zero-flux conditions are imposed on all boundaries, except at the substrate, where a static contact-angle condition is prescribed to account for droplet–substrate wetting [31].

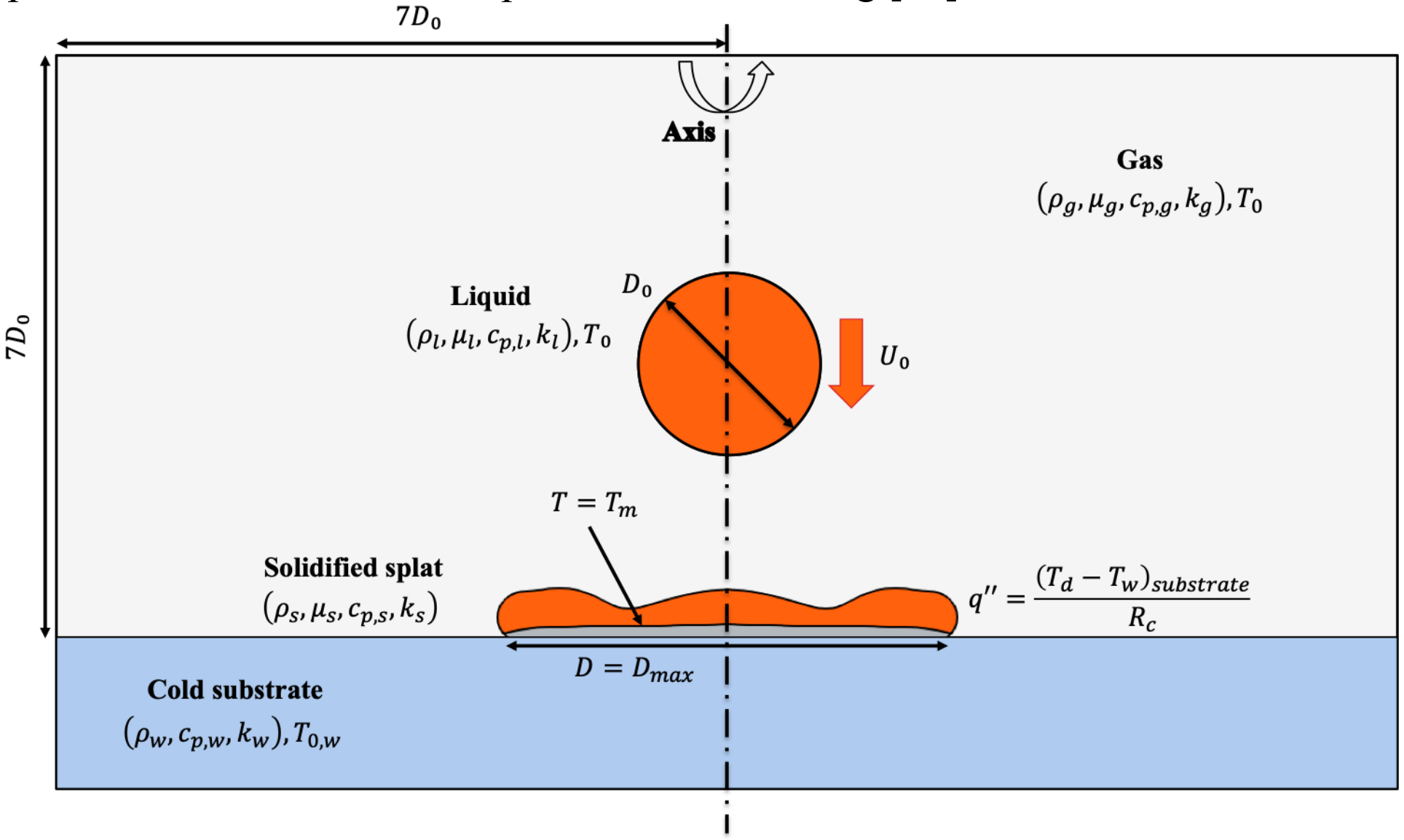

**Fig. 3.** Schematic of the axisymmetric solidifying normal impact problem for a molten droplet in a $7D_0 \times 7D_0$ droplet-fluid domain, including the initial droplet, cold substrate, maximum spread diameter, solidified splat, and thermal contact resistance at the wetted interface.

### 3.3. *Mesh adaptivity*

Adaptive mesh refinement (AMR) is used to efficiently resolve the localized gradients that control the impact and solidification dynamics. Three refinement indicators are employed: $\nabla c$ to capture the diffuse gas–condensed-phase interface, $\nabla f_l$ to resolve the mushy region where the liquid fraction varies rapidly, and near-wall refinement along the cold substrate to capture thermal gradients and hydrodynamic boundary-layer effects. This strategy concentrates resolution near the moving interface, solidification front, and wall regions, while allowing coarser elements away from the droplet-substrate interaction zone. A representative AMR snapshot is shown in Fig. 4, where the inset views illustrate the refined mesh around the gas–condensed-phase interface, the thermal solidification front, and the melt–solidified-droplet–gas triple-line region.

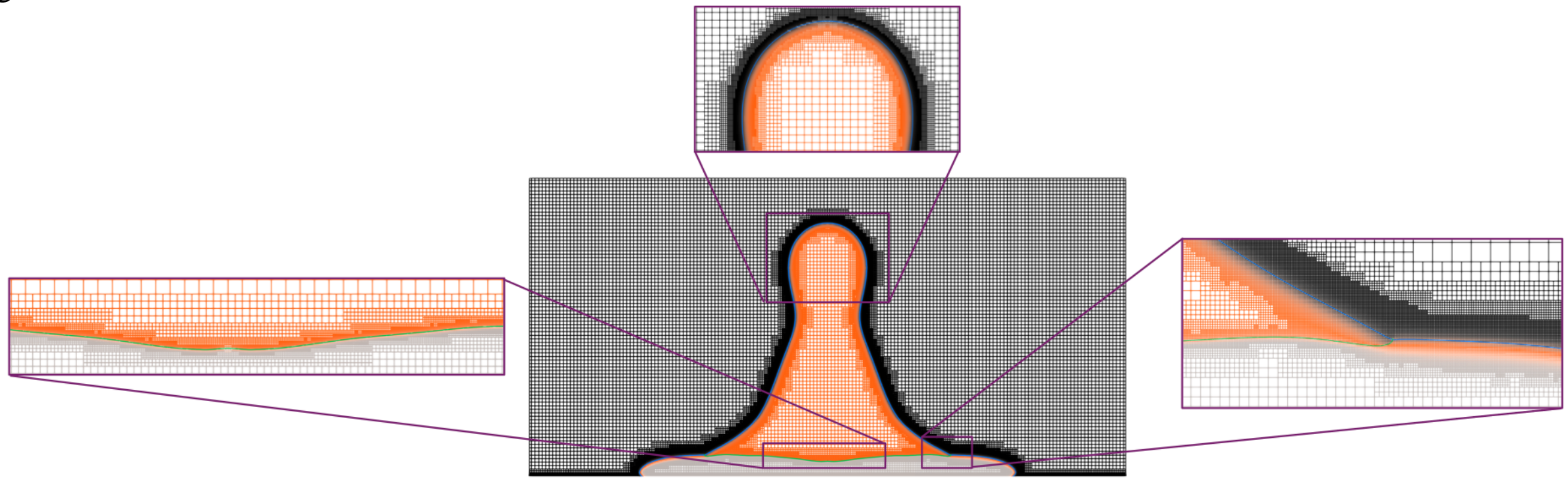

**Fig. 4.** Representative adaptive mesh during solidifying normal impact of a molten droplet. The main panel presents the mesh distribution around the deforming droplet, while the insets highlight targeted refinement near the gas–condensed-phase interface, the solidification-front region, and the liquid–solidified-droplet–gas triple-line region. The blue contour denotes $c = 0$, corresponding to the sharp diffuse gas–condensed interface, while the green contour denotes $T = T_m$, corresponding to the thermal solidification front.

## 4. Results and discussion

In this section, three cases are presented. First, the benchmak one-dimensional Stefan problem is used to validate the phase-change heat-transfer model. Next, the isothermal normal impact of a molten tin droplet is simulated and compared with the experiments of Aziz and Chandra [1]. Finally, the normal impact of a solidifying molten tin droplet is simulated using the proposed free-energy modifications, and the results are compared against the experiments of Aziz and Chandra [1] and against other numerical and analytical predictions.

The key dimensionless groups for solidifying droplet impact are

$$Re = \frac{\rho U_0 D_0}{\mu}, We = \frac{\rho U_0^2 D_0}{\gamma}, Ste = \frac{c_p (T_m - T_w)}{L}, Pr = \frac{\mu c_p}{k}, \tag{26}$$

where $Re$ and $We$ describe the impact hydrodynamics, while $Pr$ characterizes momentum diffusion relative to thermal diffusion. The Stefan number $Ste$ measures the relative magnitude of the charchteristic internal energy to latent heat and therefore characterizes the thermal capacity of the system to drive the solidification process.

Properties of molten and solidified tin, the stainless-steel substrate, and the surrounding gas used in the numerical simulations are listed in Table 1. The oxidation effects are neglected, and the surface tension is assumed to remain constant. A finite high effective viscosity is assigned to solidified tin to suppress motion in the solidified region.

**Table 1**. Thermophysical properties of molten tin, solidified tin, air, and stainless steel used in the numerical simulations.

| Phase | $\rho$ $(\mathrm{kg \cdot m^{-3}})$ | $\mu$ $(\mathrm{Pa \cdot s})$ | $c_p$ $(\mathrm{J \cdot kg^{-1} \cdot K^{-1}})$ | $k$ $(\mathrm{W \cdot m^{-1} \cdot K^{-1}})$ | $L$ $(\mathrm{J \cdot kg^{-1}})$ | $\gamma$ $(\mathrm{N \cdot m^{-1}})$ |
|---|---|---|---|---|---|---|
| Liquid tin | 6979 | $1.90 \times 10^{-3}$ | 250 | 32.0 | 59000 | 0.560 |
| Solid tin | 7200 | $\approx 1.90 \times 10^{1}$ | 216 | 60.0 | ~ | ~ |
| Air | 1.18 | $1.85 \times 10^{-5}$ | 1006 | $2.63 \times 10^{-2}$ | ~ | ~ |
| Stainless steel | 7900 | ~ | 477 | 14.9 | ~ | ~ |

### 4.1. *Stefan problem*

For validation of the phase-change heat-transfer model, solidification in a one-dimensional semi-infinite medium is considered, as sketched in Fig. 5. The domain is initially filled with liquid at temperature $T_0 = 1$. At $t = 0$, the left wall is suddenly cooled below the melting temperature to the subcooled value $T_w = -1$, while the melting temperature is $T_m = 0$.

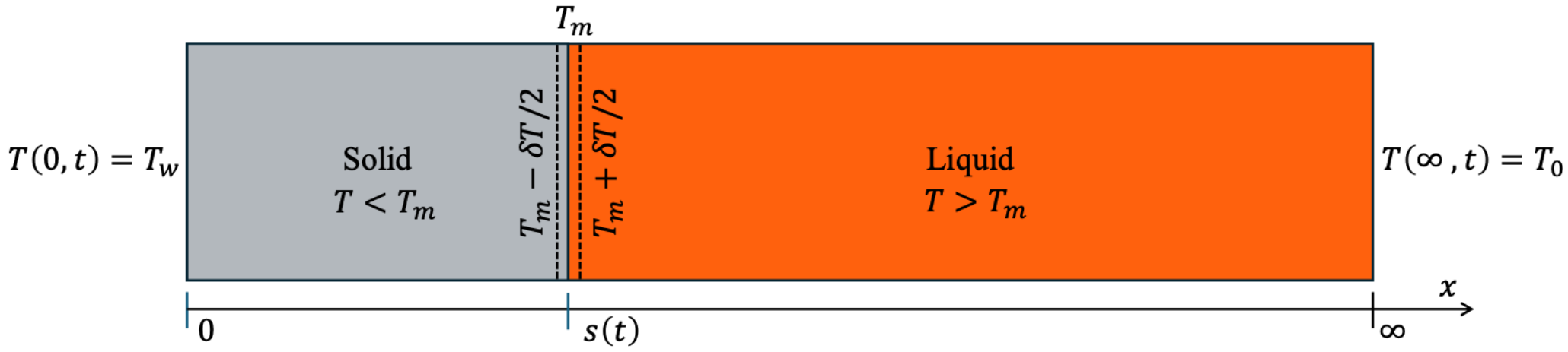

**Fig. 5.** Schematic of the one-dimensional Stefan problem used for validation of the phase-change heat-transfer model. The liquid is initially at $T_0$, the left wall is maintained at $T_w < T_m$, and the solid–liquid

interface $s(t)$ advances into the liquid. The phase change is regularized over a narrow temperature interval around $T_m$.

Freezing (solidification) begins at the cold wall, and the solid–liquid interface propagates into the liquid phase. The solidification-front position for the two-phase Stefan problem is obtained from the self-similar Neumann solution [32],

$$s(t) = 2\lambda\sqrt{\alpha_s t}, \tag{27}$$

where $s(t)$ is the solidification-front location, $\alpha_s = k_s/(\rho_s c_{p,s})$ is the thermal diffusivity of the solid phase, and $\lambda$ is obtained from [32]

$$\frac{Ste_s}{\exp(\lambda^2)\mathrm{erf}(\lambda)} - \frac{Ste_l\sqrt{\alpha_l/\alpha_s}}{\exp(\lambda^2\alpha_s/\alpha_l)\mathrm{erfc}\left(\lambda\sqrt{\alpha_s/\alpha_l}\right)} = \lambda\sqrt{\pi}. \tag{28}$$

Here, $Ste_s = c_{p,s}(T_m - T_w)/L$ and $Ste_l = c_{p,l}(T_0 - T_m)/L$ are the Stefan numbers of the solid and liquid phases, respectively. The analytical temperature field is then given by [32–34]

$$T(x,t) = \begin{cases} T_w + (T_m - T_w)\dfrac{\mathrm{erf}\left(\dfrac{x}{2\sqrt{\alpha_s t}}\right)}{\mathrm{erf}(\lambda)}, & 0 \le x \le s(t), \\ T_0 + (T_m - T_0)\dfrac{\mathrm{erfc}\left(\dfrac{x}{2\sqrt{\alpha_l t}}\right)}{\mathrm{erfc}\left(\lambda\sqrt{\dfrac{\alpha_s}{\alpha_l}}\right)}, & x > s(t). \end{cases} \tag{29}$$

The analytical solution is used to validate both the predicted temperature field and the evolution of the solidification front. For this benchmark validation case, the parameters are set to $Ste_l = 0.010$, $Ste_s = 0.005$, $\alpha_l = 0.04$, $\alpha_s = 0.36$, and $\delta T = 0.02$. A sufficiently large computational domain is used to approximate the semi-infinite medium.

The plots in Fig. 6 compare the numerical predictions with the analytical solutions for the solidification-front position and temperature profiles at selected time instants. The close agreement between the numerical and analytical results verifies that the model accurately captures the coupled conduction and latent-heat effects governing the solidification process. This validates the enthalpy-based treatment of phase change, in which latent heat is incorporated as a volumetric source term in Eq. (1).

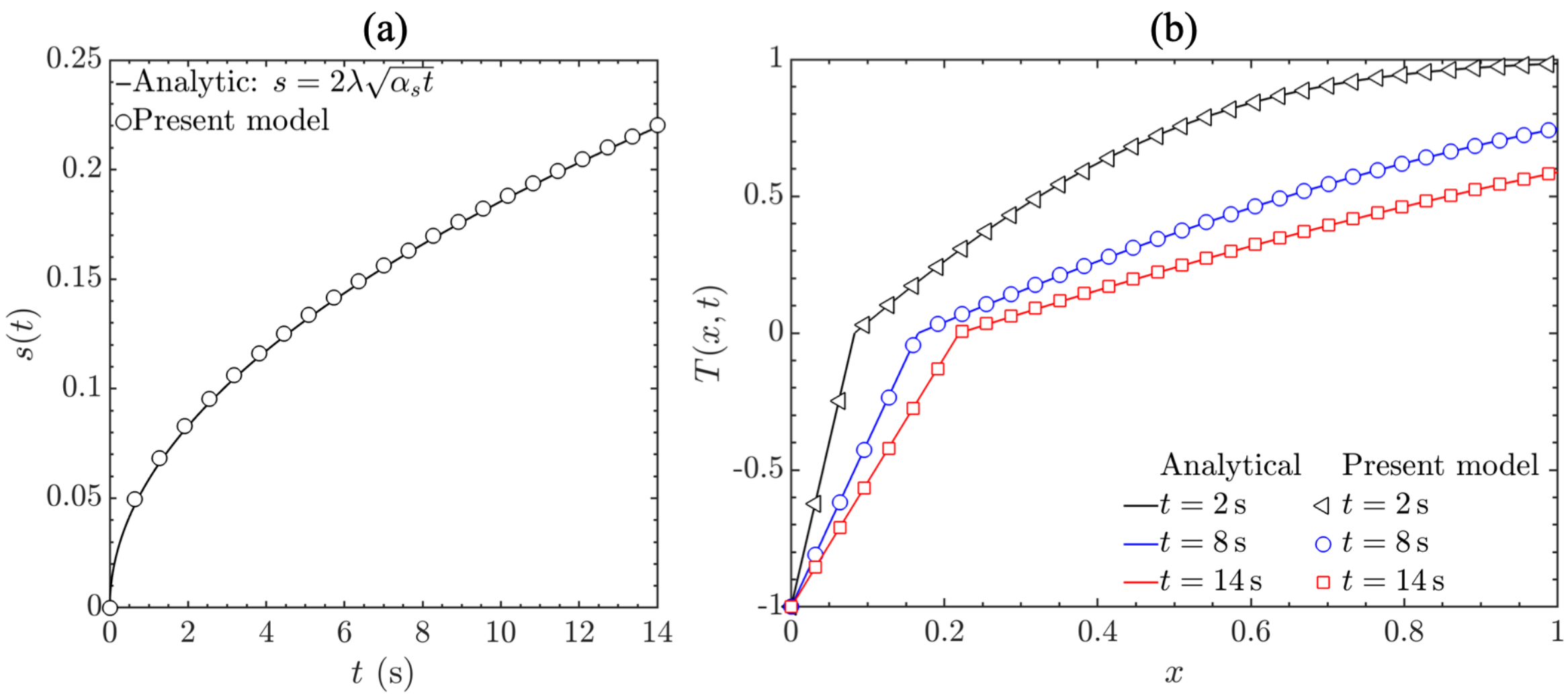

**Fig. 6.** Validation of the phase-change heat-transfer model using the one-dimensional Stefan problem: (a) temporal evolution of the solidification-front position and (b) temperature profiles at selected time moments. Numerical results from the present model are compared with the analytical self-similar solution for a semi-infinite medium.

### 4.2. *Isothermal normal drop impact*

The isothermal normal impact of a molten tin droplet is numerically simulated using the properties listed in Table 1. Following the experiments of Aziz and Chandra [1], the droplet is maintained at $T = 240°\mathrm{C}$, slightly above the melting temperature of tin, $T_m = 232°\mathrm{C}$. Under these conditions, solidification is not expected during impact, and the problem is treated as isothermal.

The static contact angle between the molten tin droplet and the stainless-steel substrate is prescribed as $140°$, consistent with the experimental measurements [1]. The initial droplet diameter is $D_0 = 2.7$ mm, and the impact velocity is $U_0 = 1\ \mathrm{m/s}$, corresponding to $Re \approx 9900$ and $We \approx 34$. The diffuse-interface thickness is set to $\xi = D_0/120$.

The isothermal impact case is used as a baseline validation of the hydrodynamic and capillary components of the CHNS model before introducing solidification. The results in Fig. 7 compare the predicted droplet shapes with the experimental snapshots of Aziz and Chandra [1]. The present model captures the main stages of impact, including the initial spreading, maximum spread, recoil, rebound behavior, and vertical jet formation. This agreement confirms that the phase-field formulation reproduces the essential isothermal impact dynamics of molten tin droplets.

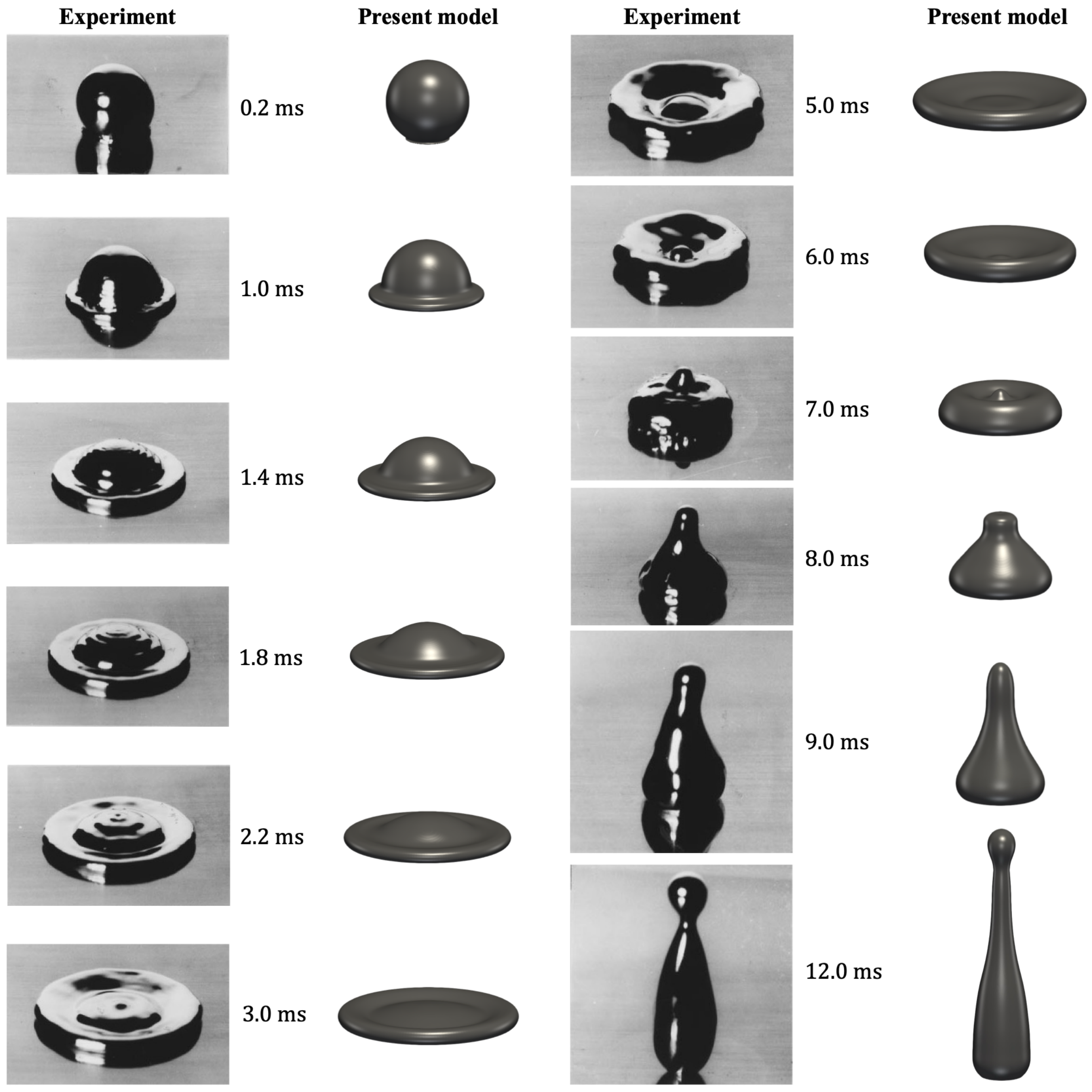

**Fig. 7.** Comparison of experimental snapshots [1] and present phase-field predictions for the isothermal normal impact of a molten tin droplet on a stainless-steel substrate at $U_0 = 1\ \mathrm{m/s}$. The sequence reveals spreading, maximum deformation, recoil, and vertical jet formation at selected time moments.

The spread factor is defined as $\zeta = D(t)/D_0$, where $D(t)$ is the instantaneous wetted diameter and $D_0$ is the initial droplet diameter. The predicted spread-factor evolution is compared with experimental measurements [1] and previous phase-field simulations of Shen st al. [12] in Fig. 8. The present model captures the rapid initial spreading, maximum spread, and subsequent recoil of the molten tin droplet. In particular, the present model predicts a maximum spread factor of $\zeta_{\max} = 2.12$, in close agreement with the experimental value of $\zeta_{\max} = 2.20$, with a relative difference of approximately 3.6%. This provides an accurate prediction of the maximum spreading while preserving good agreement with the overall temporal evolution.

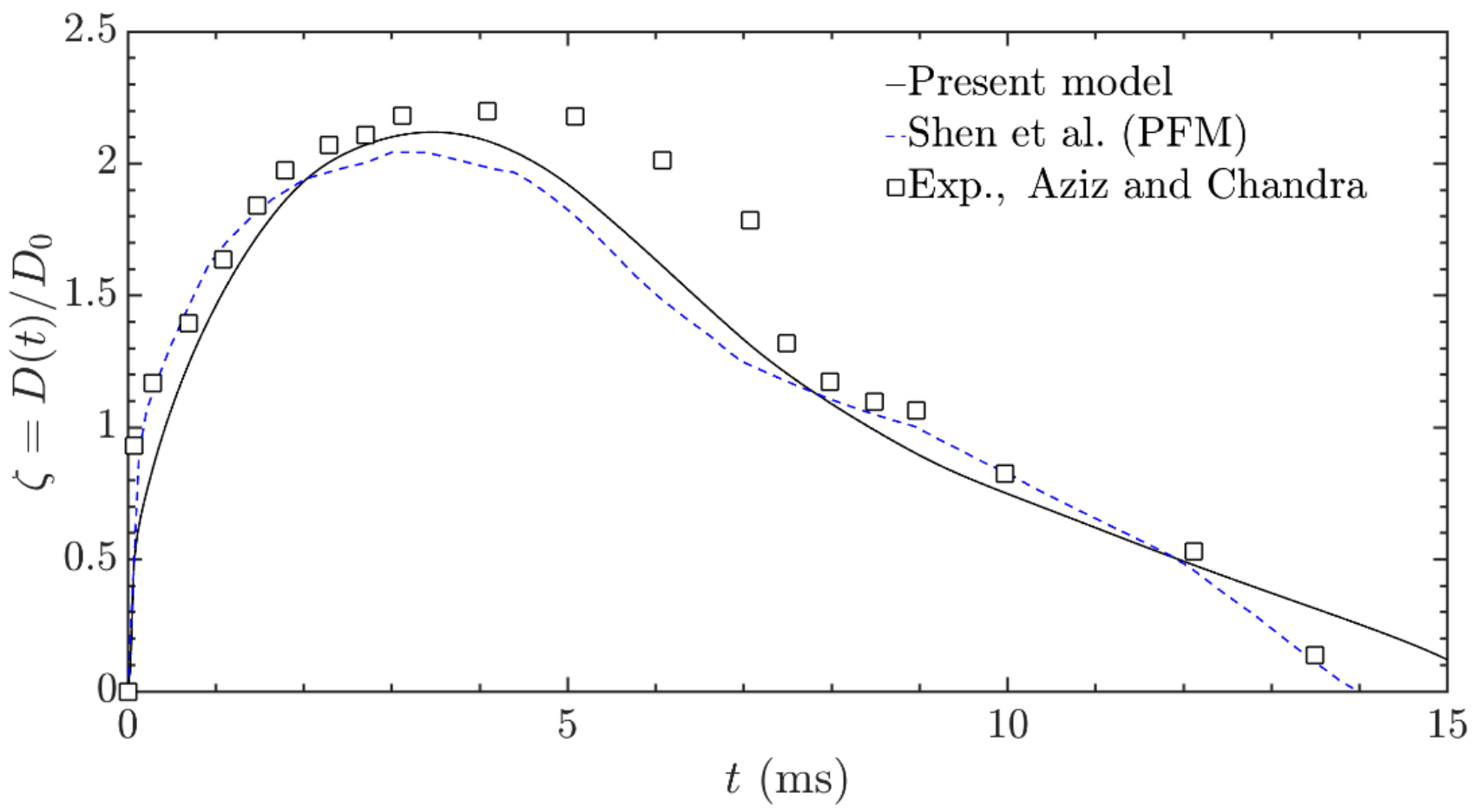

**Fig. 8**. Temporal evolution of the spread factor for isothermal normal impact of a molten tin droplet. Present phase-field predictions are compared with the experiments [1] and other phase-field results [12].

4.3. *Drop impact with solidification*

Having demonstrated the accuracy of the phase-change heat-transfer model and the isothermal impact dynamics, the normal impact of a solidifying molten tin droplet is simulated using the same droplet diameter, impact velocity, static contact angle, and the interfacial thickness as in the isothermal case. A constant phase-field mobility, $M = 10^{-9}\ \mathrm{m}^2/(\mathrm{Pa\ s})$ is adopted.

All thermophysical properties used in the simulations are listed in Table 1. The substrate is initially at $T_{0,w} = 25°\mathrm{C}$, while the pre-impact molten droplet temperature is maintained at $T_0 = 240°\mathrm{C}$. Consistent with the VOF simulations of Pasandideh-Fard et al. [6], a constant thermal contact resistance of $R_c = 5 \times 10^{-6}\ \mathrm{m}^2\ \mathrm{K/W}$ is prescribed over the wetted droplet–substrate footprint. The melting temperature of tin is taken as $T_m = 232°\mathrm{C}$, and the solid–liquid transition is regularized using a temperature offset of $\delta T = 2°\mathrm{C}$. The low Prandtl number of molten tin, $Pr = \mathcal{O}(10^{-2})$, indicates that thermal diffusion dominates momentum diffusion during impact, promoting rapid cooling and solidification.

For the solidifying-impact simulations, the internal wall-energy coefficient and solidification penalty strength are set to $\beta_w = -0.54$ and $\mathrm{K}_0 = 10^5\ \mathrm{J/m}^3$, respectively. The coefficient $\beta_w$ is treated as an effective calibration parameter, similar in role to the wall-energy parameters used in the phase-field contact-angle formulations [13]. Several values are tested, and $\beta_w = -0.54$ is selected because it yields the best agreement between the predicted splat morphology and the experiments. With the present sign convention, negative values of $\beta_w$ promote recoil of the remaining melt from the solidified region. The parameter $\mathrm{K}_0$ controls the strength of the solidification penalty. It is chosen large enough to anchor the phase field in the solidified region, but not so large that the penalty over-constrains the diffuse interface, introduces excessive penalty-induced diffusion, or artificially alters the drop morphology.

A baseline $200 \times 200$ mesh with four levels of adaptive refinement is used to resolve the gas–condensed-phase interface, the thermal solidification front, and the near-substrate boundary-layer region. As the adaptive mesh evolves during the impact, the nonlinear system in the fluid domain contains approximately $(0.81 - 0.93) \times 10^6$ degrees of freedom. This case introduces the coupled effects of droplet deformation, heat transfer, solidification, and the proposed free-energy modifications for melt retraction and solidified-region immobilization.

The present simulations are compared with the experimental observations [1,6] in Fig. 9, revealing good qualitative agreement over the full impact sequence. The model reproduces the initial spreading stage and the maximum spread at approximately $t = 4.5$ ms. The main differences among numerical approaches become more pronounced after maximum spread, when surface-tension-driven recoil and melt retraction over the solidified splat dominate the dynamics [35].

In a standard CHNS formulation, suppressing the velocity inside the solidified region is insufficient, because the remaining melt does not experience the solidified portion as a wetting boundary. In the present formulation, the diffuse-domain wall-energy term provides this missing energetic pathway, enabling the melt to retract over its solidified portion. This behavior is evident at $t = 8.3$, 11.3, 15.3, and 17.3 ms, where the predicted morphology follows the experimentally observed recoil and retraction sequence. Unlike approaches that prescribe a contact angle for molten tin on its own solidified phase during recoil, the present model does not impose a geometric contact angle at the evolving internal solidified region. Instead, the retraction is driven energetically through the localized wall-energy term, with its strength controlled by the parameter $\beta_w$.

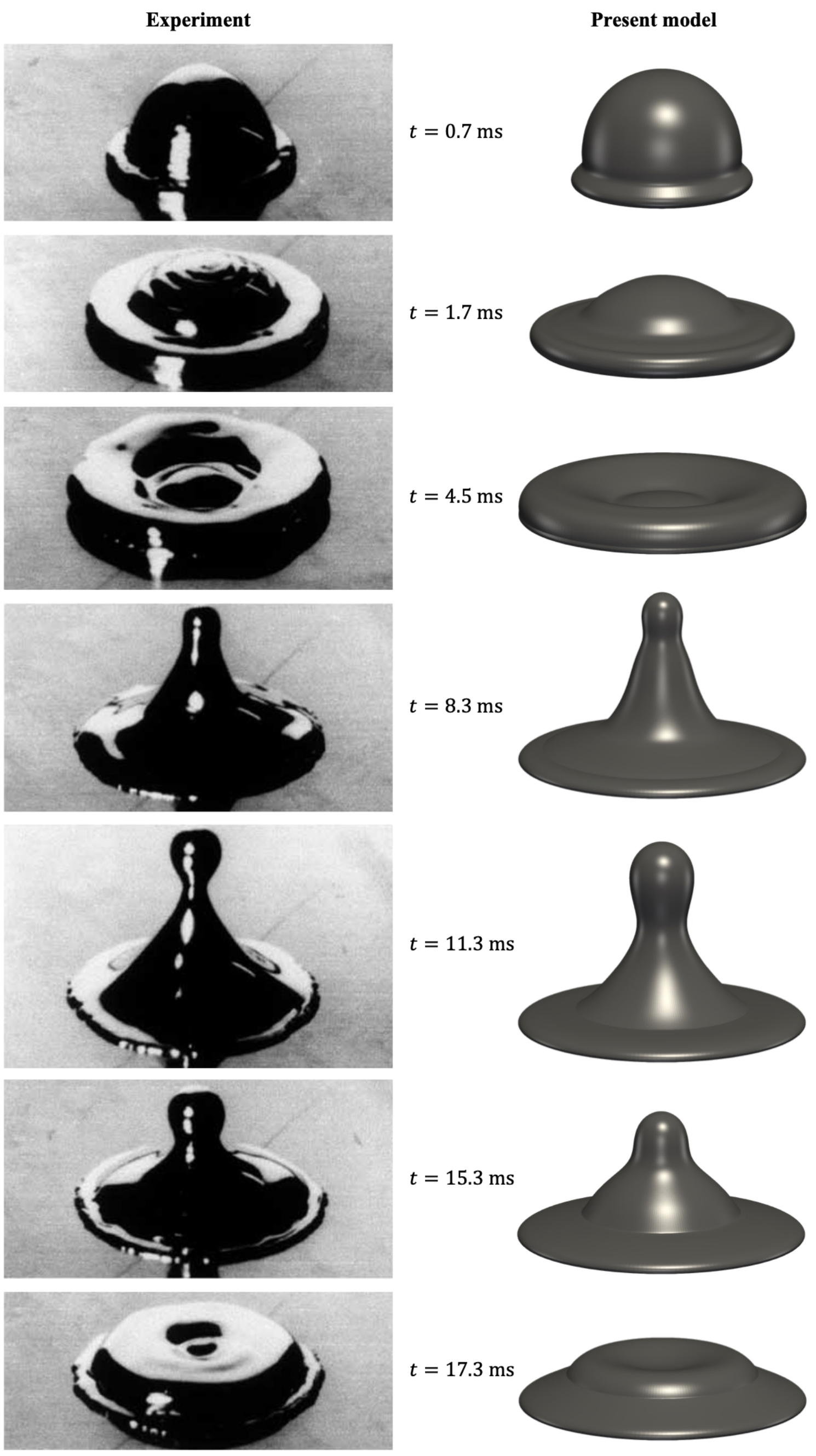

**Fig. 9.** Comparison of experimental snapshots [1,6] with the present numerical predictions for the normal impact of a solidifying molten tin droplet onto a stainless-steel substrate. The sequence reveals the initial spreading, maximum spread, recoil, and melt retraction over the solidified splat at selected times.

The temperature evolution during solidifying molten tin droplet impact is illustrated in Fig. 10. Immediately after the impact, a large temperature difference between the molten tin droplet and the initially cold substrate drives intense heat transfer across the wetted droplet-substrate contact region. As a result, the strongest cooling within the droplet occurs near the basal contact region, where solidification first initiates. Simultaneously, the substrate is heated locally beneath the wetted footprint, forming a confined thermal penetration region that expands with the spreading droplet. As spreading proceeds, the basal solidified layer grows along the droplet-substrate contact interface and contributes to the contact-line arrest near the maximum-spread stage at $t \approx 4.5$ ms. After this stage, the solidification front continues to thicken near the substrate while the upper portion of the melt remains above $T_m$, leading to a distinct separation between the immobilized solidified splat and the retracting liquid core.

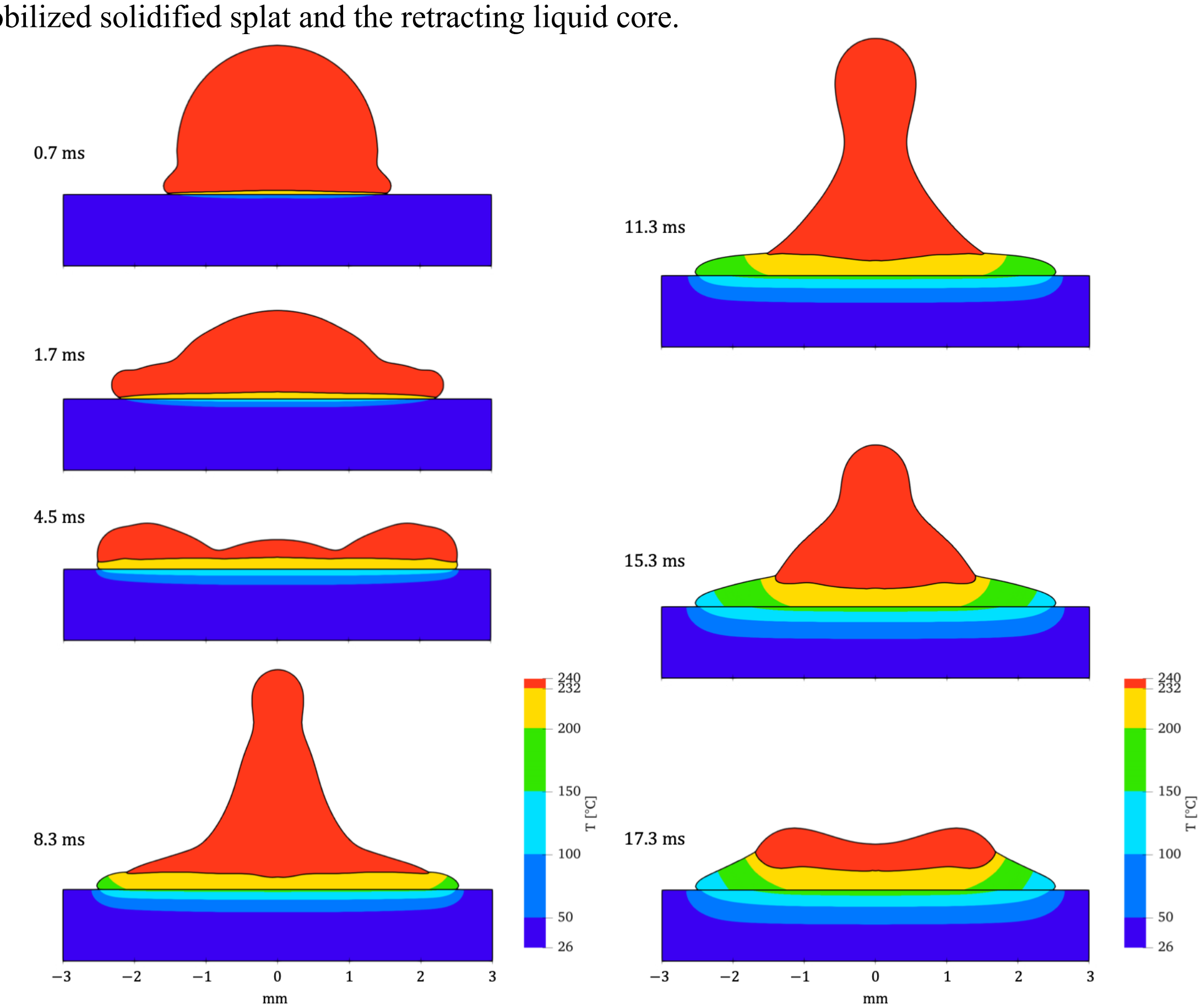

**Fig. 10.** Temporal evolution of a solidifying molten tin droplet impact. The panels display droplet morphology and temperature distribution in both the droplet and substrate at selected time moments. The black solid lines denote the gas–condensed-phase interface obtained from the order parameter and the solidification front. The contours illustrate heat extraction into the cold substrate, basal solidification, melt recoil, and retraction over the solidified splat after maximum spreading.

The local thermal response is further quantified in Fig. 11(a) using the substrate-side temperature $T_w$, and the interfacial heat flux $q'' = (T_d - T_w)/R_c$ sampled at the center of the wetted interface. The heat flux is initially high because of the significant temperature jump across the thermal contact resistance, but it decreases rapidly as the substrate warms and the droplet cools near the interface. At later times, continued heat transfer into the substrate promotes further growth of the solidified region, while the remaining liquid recoils and retracts over the solidified splat.

Yarin et al. [3] predict the solidification height using a Stefan-type self-similarity solution. Assuming perfect thermal contact between the splat and the substrate, the solidification front is located at a constant similarity coordinate $\Xi^*$, yielding $Z_s = \Xi^* \sqrt{\nu_l t}$, where $\nu_l$ is liquid kinematic viscosity. The model couples thermal balance to the self-similar viscous flow in the spreading lamella, while assuming one-dimensional heat transfer normal to the wall, semi-infinite liquid and substrate bodies, and constant thermophysical properties. Augmenting their approach to include thermal contact resistance is straightforward. The temperature jump at the contact between the splat and the substrate makes the solid-side temperature-gradient factor time-dependent, and therefore $\Xi^*$ becomes time-dependent too.

The results in Fig. 11(b) compare the solidification height predicted by the extended Yarin et al. model [3] with time-dependent $\Xi^*$, with the present phase-field results. The solidification-front height profiles in Fig. 11(c) reveal that the solidified splat thickness is spatially nonuniform. Therefore, for the phase-field results in Fig. 11(b), the solidification-front height is reported as a diameter-averaged value over the radial extent of the extracted $T = T_m$ front. The numerical prediction follows the analytical solution well throughout the full spreading phase. The deviation becomes noticeable after $t \approx 12$ ms, when the numerical result exhibits a reduced growth rate. This late-time behavior coincides with droplet recoil and the development of a radially nonuniform molten region, while the analytical model assumes a one-dimensional, semi-infinite liquid layer above the solidified splat. Thus, the later discrepancy may reflect the effect of recoil-induced morphology on the diameter-averaged front height.

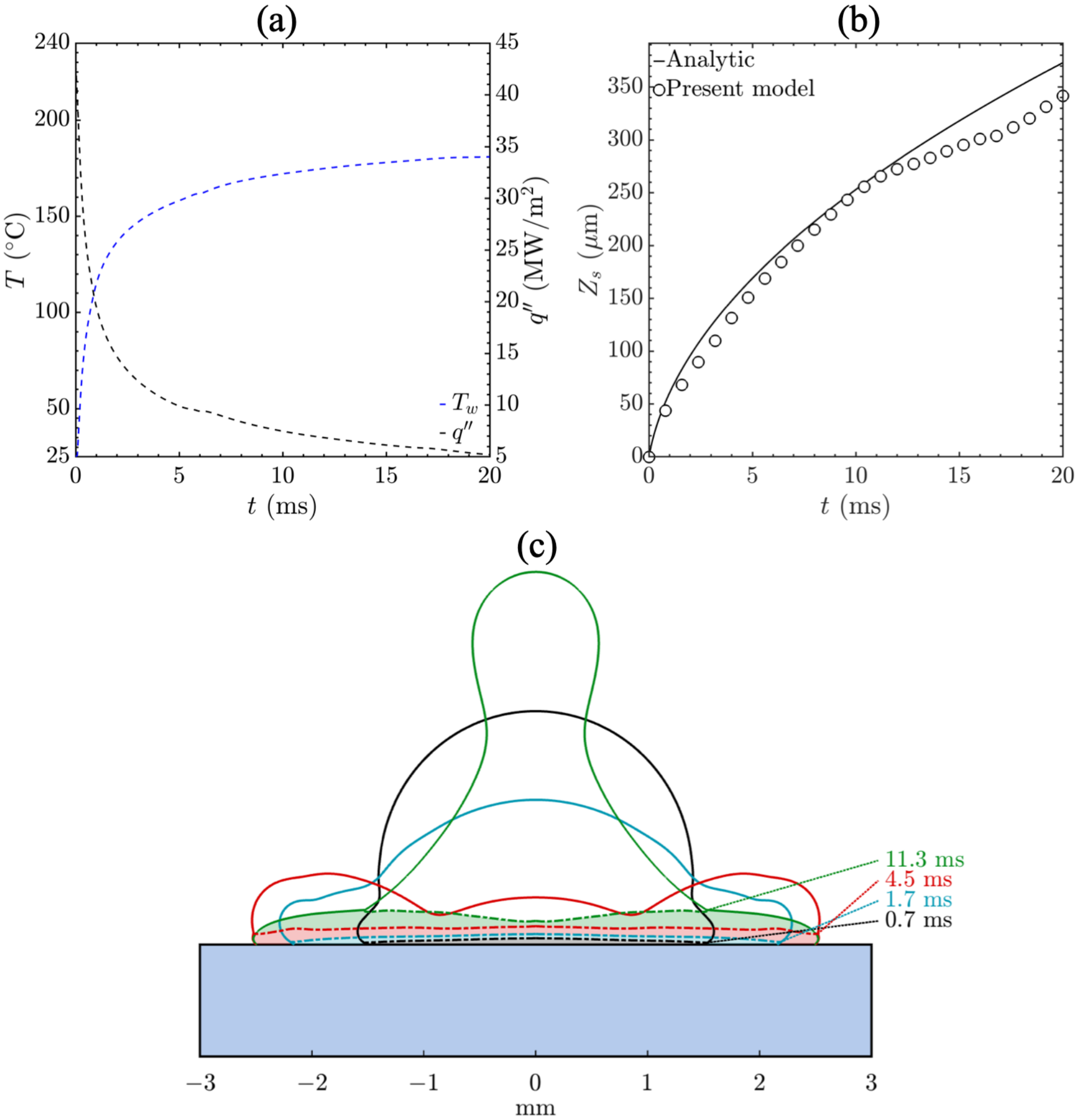

**Fig. 11.** Temporal evolution of (a) the substrate-side temperature $T_w$, and interfacial heat flux $q''$at the center of the wetted interface, (b) solidification height $Z_s$ predicted by the extended analytical model [3] and the present phase-field results, and (c) droplet and solidification-front profiles at selected moments. In panel (c), the shaded regions correspond to newly solidified regions, the solid lines represent the droplet profiles, and the dash-dotted lines indicate the solidification-front profiles.

To isolate the role of the wall-energy term $f_w$, Fig. 12 compares post-maximum-spread morphologies obtained with $\beta_w = -0.54$ and $\beta_w = 0$, while keeping the same solidification penalty treatment. When $\beta_w = 0$, the internal wall-energy contribution is inactive, and the formulation behaves like a classical single-order-parameter CHNS model. In this case, the remaining melt stays attached to the solidified splat throughout the post-spreading stage, producing a morphology similar to previous predictions based on the classical CHNS, such as those of Shen et al [12]. This occurs because the liquid and solidified tin are represented by the same condensed-phase order parameter, so the baseline free energy does not distinguish liquid-on-solid wetting from liquid–gas interfacial energy and therefore provides no energetic driving mechanism for melt detachment from the frozen splat. In contrast, $\beta_w = -0.54$ introduces an effective dewetting bias localized near the diffuse melt–solidified-splat–gas triple-line region. This increases the energetic cost of maintaining melt contact with the solidified region and enables the remaining liquid to recoil over the immobilized splat. The comparisons demonstrate that the proposed wall-energy correction supplies the missing energetic mechanism required to capture the recoil of the retracting liquid core over the solidified splat.

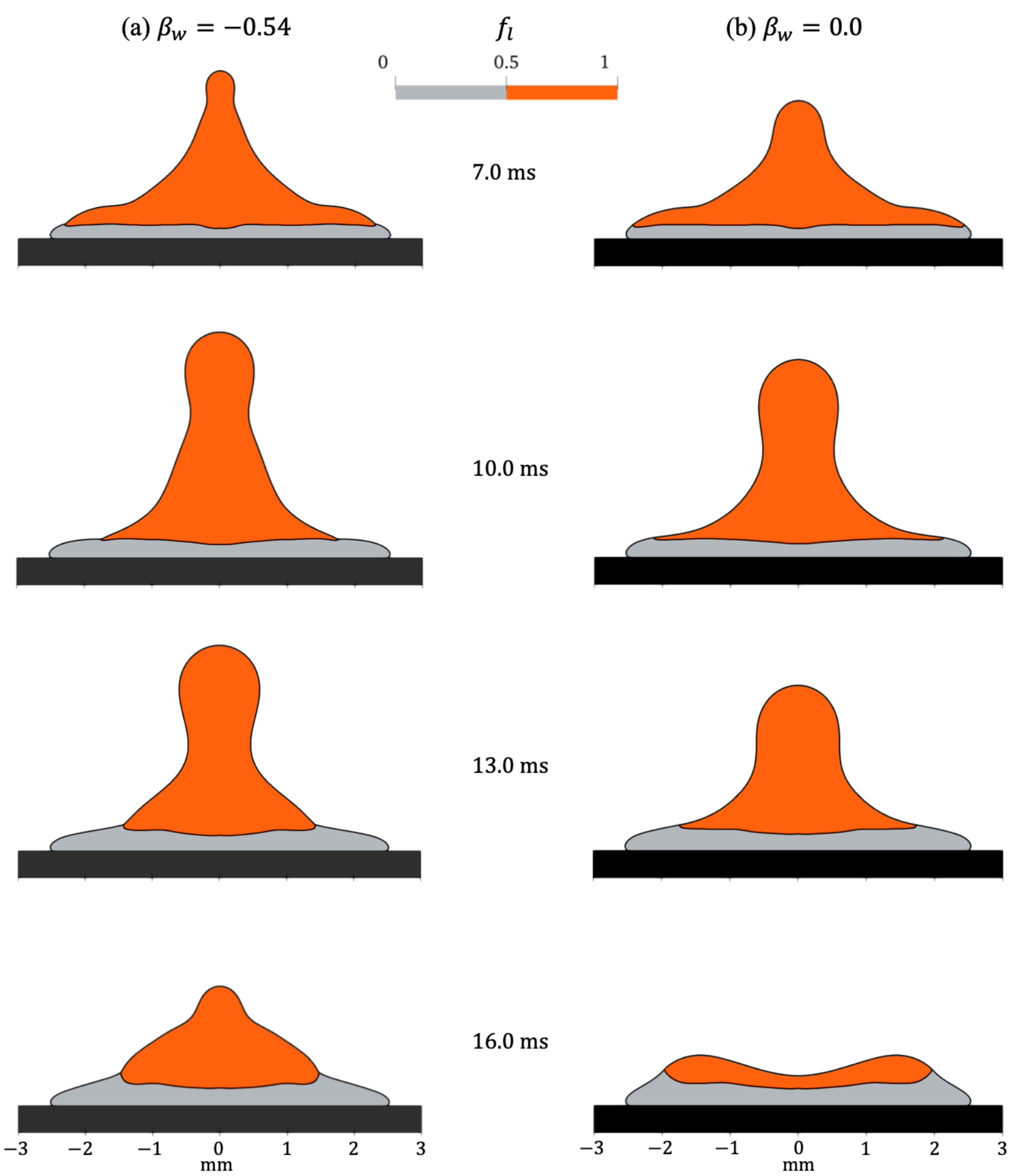

**Fig. 12.** Effect of the diffuse-domain wall-energy correction during post-maximum-spread melt retraction. Snapshots compare the liquid-fraction field for: (a) $\beta_w = -0.54$ and (b) $\beta_w = 0.0$. The color field denotes the liquid fraction, where $f_l = 0$ corresponds to solidified tin, and $f_l = 1$ corresponds to liquid tin.

The predicted spread factor is compared with experiments and previous numerical results in Fig. 13. The present phase-field model and the VOF simulations of Pasandideh-Fard et al. [6] use the same thermal contact resistance, $R_c = 5 \times 10^{-6}$ m$^2$K/W, whereas the previous phase-field simulation of Shen et al. [12] used a higher value, $R_c = 1 \times 10^{-5}$ m$^2$K/W. The higher thermal contact resistance reduces heat extraction into the substrate, delays solidification, and allows a larger maximum spread. Therefore, differences in the prescribed thermal contact resistance should be taken into account when interpreting the predicted maximum spreading. Importantly, the late-stage stabilized spread factor is predicted more accurately than in the previous numerical results. The present model yields $\zeta \approx 1.86$, which is closer to the experimental plateau. Together with the splat morphology comparisons, this demonstrates that the proposed numerical framework better captures the coupled spreading, solidification, and post-maximum-spread retraction dynamics of solidifying molten tin droplet impact.

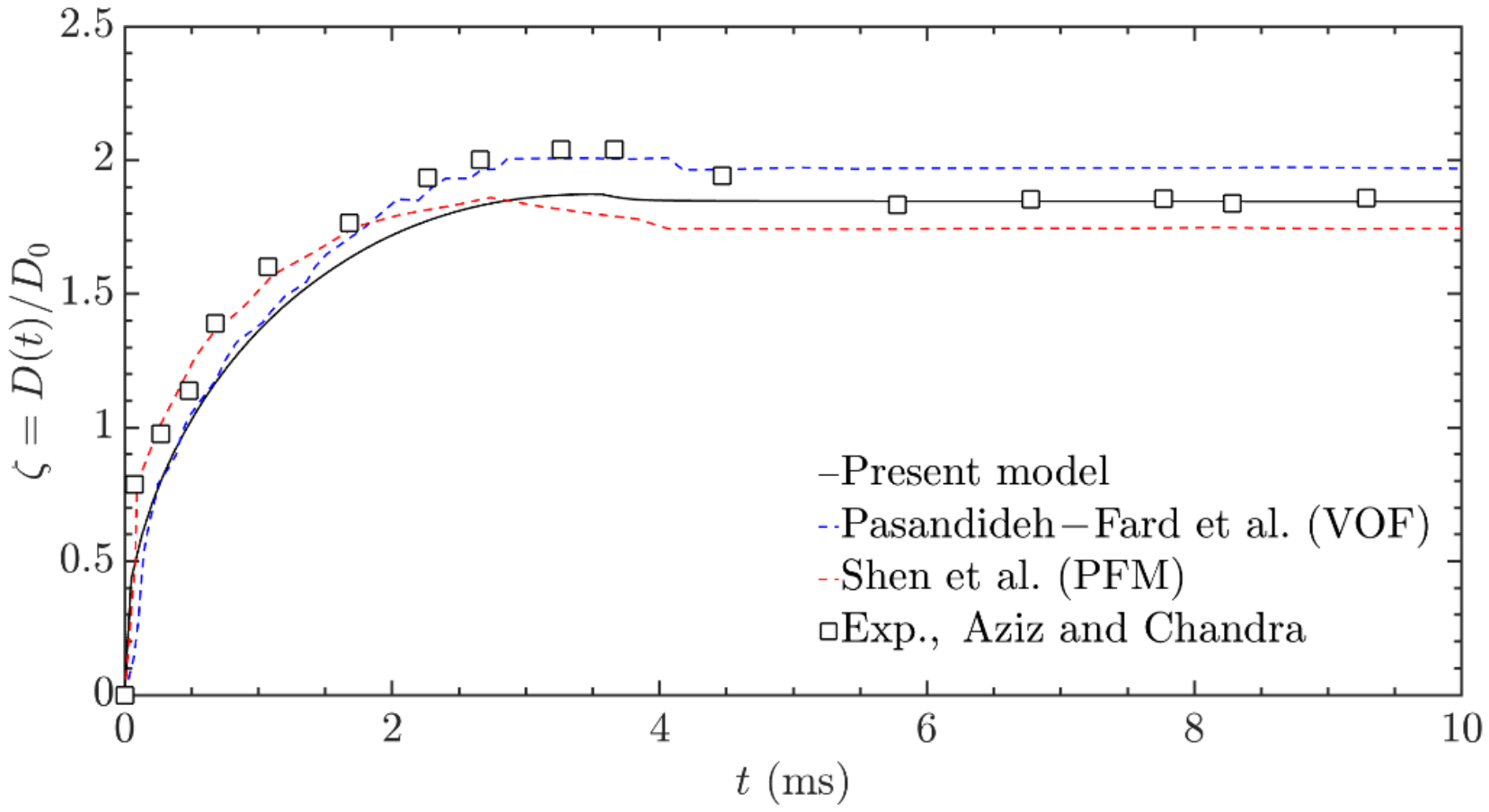

**Fig. 13.** Temporal evolution of the spread factor for the normal impact of a solidifying molten tin droplet. Present predictions are compared with the experiments of Aziz and Chandra [1], the VOF results of Pasandideh-Fard et al. [6], and the phase-field results of Shen et al. [12].

## 5. Summary and conclusions

This study presents an improved phase-field framework for simulating the impact and solidification dynamics of molten metal droplets in the high-Weber-number regime. A key limitation of the conventional CHNS formulation is that the solidified portion of the droplet is not treated as an internal wetting boundary. As a result, the remaining melt lacks an energetic pathway to recoil over the frozen splat after maximum spreading. To address this limitation, a diffuse-domain wall-energy term was introduced, with its chemical-potential contribution localized near the liquid–solidified-splat–gas triple-line region. This term provides an effective wetting/dewetting mechanism for the remaining melt without prescribing a geometric contact angle on the evolving internal solidified region.

A solidification penalty term is also incorporated into the free-energy functional to suppress artificial motion of the gas–solid interface caused by residual Cahn–Hilliard diffusion. This term anchors the phase-field variable in frozen regions while avoiding ad hoc treatments such as manually pinning the contact line after maximum spreading. The formulation is implemented in a fully coupled finite-element framework with adaptive mesh refinement to resolve the gas–droplet interface, solidification front, and near-wall thermal and hydrodynamic gradients.

The model is first validated using the one-dimensional Stefan problem and normal impact of an isothermal droplet. It is then applied to the normal impact of a solidifying molten tin droplet onto a stainless-steel substrate. The proposed formulation reproduces the post-maximum-spread recoil of the remaining melt over the solidified splat and predicts the stabilized final splat diameter with reasonable accuracy. The comparison with the baseline CHNS formulation reveals that without the diffuse-domain wall-energy correction, the remaining melt stays attached to the frozen splat, similarly to previous predictions based on the classical CHNS. Physically, the improvement arises because the added wall-energy term allows the solidified region to act as an effective internal wetting/dewetting surface for the remaining melt, rather than treating liquid and solidified tin as the same condensed phase with no interfacial energetic preference.

The present formulation introduces one additional parameter, $\beta_w$, which controls the localized wetting/dewetting bias near the diffuse triple-line region. Physically, $\beta_w$ represents the effective interaction between the remaining melt and its own frozen splat, including the effect of melt-on-solidified-splat wettability. In the present model, it is treated as a tunable parameter, and the calibrated value adopted here enables the model to reproduce the experimentally observed post-maximum-spread droplet profiles in which the remaining melt recoils over the immobilized solidified region.

**Acknowledgements**

The authors acknowledge the financial support from the National Science Foundation award CBET 2312197. Simulations were performed using the High-Performance Computing (HPC) resources supported by the University of Arizona TRIF, UITS, and Research, Innovation, and Impact (RII) and maintained by the University of Arizona Research Technologies team.